\documentclass[aps,prb,twocolumn,amsmath, superscriptaddress, floatfix, showpacs]{revtex4}
\usepackage{graphicx}

\newcommand{\bv}[1]{{\bf #1}}

\newcommand{\ad} {a^\dagger}
\newcommand{\bd} {b^\dagger}
\newcommand{\cd} {c^\dagger}
\newcommand{\dd} {d^\dagger}

\begin{document}

\title{Monte Carlo simulations of ${\rm Rb_2MnF_4}$, a classical Heisenberg antiferromagnet in two-dimensions with dipolar interaction}
\author{Chenggang Zhou}

\affiliation{Center for Nanophase Materials Sciences, Oak Ridge
National Laboratory, P. O. Box 2008, Oak Ridge Tennessee, 37831-6493
USA}

\author{D. P. Landau}
\affiliation{Center for Simulational Physics, University of Georgia,
Athens Georgia, 30602 USA}

\author{T. C. Schulthess}
\affiliation {Center for Nanophase Materials Sciences, Oak Ridge
National Laboratory, P. O. Box 2008, Oak Ridge Tennessee, 37831-6493
USA}

\date{\today}

\begin{abstract}

We study the phase diagram of a quasi-two dimensional magnetic system ${\rm Rb_2MnF_4}$ with Monte Carlo simulations of a classical
Heisenberg spin Hamiltonian which includes the dipolar interactions between  ${\rm Mn}^{2+}$
spins. Our simulations reveal an Ising-like antiferromagnetic phase at low magnetic fields
and an XY phase at high magnetic fields. The boundary between Ising and XY phases is analyzed
with a recently proposed finite size scaling technique and found to be consistent with a
bicritical point at $T=0$. We discuss  the computational techniques used to handle the weak dipolar
interaction and the difference between our phase diagram
and the experimental results.

\end{abstract}
\maketitle

\section{Introduction }

The phase diagram of anisotropic Heisenberg antiferromagnets has
been studied with renormalization group (RG)
methods\cite{Kosterlitz76, Nelson75, Pelcovits76, Nelson76,
Nelson77} and Monte Carlo
simulations.\cite{Landau78,Landau81,Leidl05, Zhou06} In three
dimensions, RG calculations for $4-\epsilon$ dimensions and Monte
Carlo simulations have found an Ising-like antiferromagnetic (AF)
phase at low magnetic fields and an XY phase at high fields,
separated by a 1st order spin-flop transition line. The spin-flop
transition line terminates at a bicritical point (BCP), where it
meets the phase boundary between the XY phase and the paramagnetic
(PM) phase, and the AF-PM phase boundary. In two-dimensions, due to
the Mermin-Wagner theorem,\cite{Mermin66} a BCP with $O(3)$ symmetry
has to be at zero-temperature, which was confirmed by RG
calculations in $2+\epsilon$ dimensions for the anisotropic
non-linear $\sigma$-model.\cite{Pelcovits76,Nelson77} The XY-PM
phase boundary and AF-PM phase boundary are exponentially close to
each other while the PM phase sandwiched in between narrows as
$\exp(-4\pi/T)$. On the other hand, the continuum field theory of
this model contains an infinite number of relevant perturbations
beyond the anisotropic nonlinear $\sigma$-model. Thus, it is also
valid to argue that the multicritical point may not be $O(3)$
symmetric and occurs at a finite temperature.\cite{Vicari07} One
would look for numerical evidence that distinguishes different
scenarios. However, Monte Carlo simulations have been unable to
trace the phase boundaries of the XY and AF phases to sufficiently
low temperatures, due to the exponentially large correlation length.
Recently, a novel finite size scaling analysis was used to interpret
the data from Monte Carlo simulations.\cite{Zhou06} It was found
that the apparent spin-flop transition line was actually consistent
with a zero temperature BCP. An additional continuous degeneracy in
the ground state at the spin-flop field has also been recently
discovered.\cite{Selke07} The ground state actually bears some
similarities to a tetracritical phase; thus it was argued that the
``hidden bicritical point'' might be relabeled as the ``hidden
tetracritical point."

\begin{figure}[t]
\includegraphics[width=0.95\columnwidth]{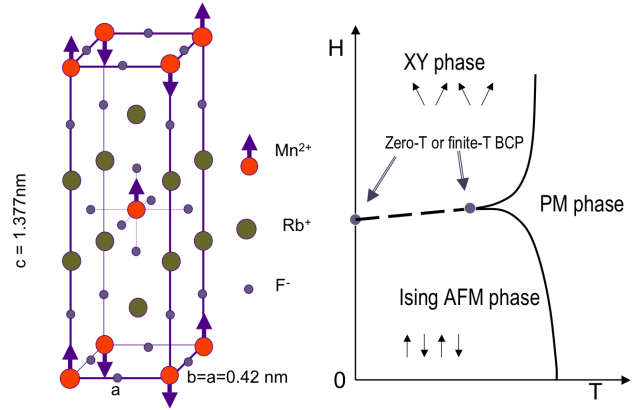}
\caption{ (color online) The unit cell of ${\rm Rb_2 Mn F_4}$ and the schematic phase diagram. If the BCP is at $T=0$, the dashed line actually represents two very close second order phase boundaries.
If the BCP is at a finite temperature, the dashed line represents a single first order phase transition. The theoretical XY phase is found to have transverse AF order in neutron scattering experiments. }
\label{fig0}
\end{figure}

In real materials, an ideal two-dimensional Heisenberg spin system has not been found, since in a
three-dimensional system, the interactions between spins can never be completely restricted to two dimensions.
Nevertheless, ${\rm Rb_2 Mn F_4}$ is a very good quasi-two-dimensional Heisenberg antiferromagnet.
In this layered compound, ${\rm Mn}^{2+}$ ions with spin-5/2 reside on (001) planes, as shown in Fig.~\ref{fig0}.
Adjacent planes are widely separated by ${\rm Rb}^+$ ions, so that the exchange interactions between
magnetic ions in different planes are negligible. 
The antiferromagnetic order parameter has been accurately measured with
neutron scattering experiments,\cite{Birgeneau70} and analyzed with spin-wave theory.\cite{Wijin73}
The theoretical model with only nearest neighbor exchanges and a staggered magnetic field
accounts for the experimental data very well. 
In the right hand portion of Fig. \ref{fig0}  we show a schematic phase diagram that summarizes the prevailing theoretical alternatives and experimental data for ${\rm Rb_2MnF_4}$.
On the other hand, the large magnetic moment of ${\rm Mn}^{2+}$ ions makes it possible to
model the spins with classical vectors. Therefore, it is an excellent system to test theoretical
predictions for two-dimensional Heisenberg spin systems, given that the effective anisotropy
due to the dipolar interaction is accounted for.\cite{Christianson01}
Obviously, the dipolar interaction plays an important role in this system, as it provides the effective
anisotropy that stabilizes the low-field AF phase and could mediate a dimensional crossover from two dimensions to three dimensions in the real material. 
With the in-plane isotropic exchange interaction and the dipolar interaction, the Neel temperature
at zero-field was calculated by Monte Carlo simulations to be 39.7$\pm$0.1~K,\cite{Lee03}
slightly higher than the experimental value 38.5$\pm$1.0~K.\cite{Wijin73, Breed67}
Following the previous research,\cite{Lee03} we performed extensive Monte Carlo simulations in both
zero and non-zero magnetic fields to construct the full phase diagram and compare it with the experiments.\cite{Cowley93}  We hope to see our model reproduce the ``apparent'' BCP at approximately $T=30$K, as seen in the experiments. To determine the phase diagram
in the thermodynamic limit, we used different finite size scaling analyses for different phase boundaries.
In particular, the ``apparent'' spin flop transition has to be examined with the novel finite size scaling method developed in Ref.~\onlinecite{Zhou06}, and it is actually found to be consistent with a zero temperature BCP.

The Hamiltonian of our model reads
\begin{eqnarray}
\label{eq1}
\nonumber {\cal H} = &-& J S(S+1) \sum_{<i,j>} \bv{S}_i \cdot \bv{S}_j
     -{U S^2 \over 2} \sum_{i\neq j, \;\alpha,\beta} S_i^\alpha {\cal D}_{ij}^{\alpha\beta} S_j^\beta\\
        &-& \sum_i S g  \mu_B \bv{h} \cdot \bv{S}_i,
\end{eqnarray}
where $S=5/2$, $\bv{S}_i$ are three dimensional unit vectors,
$J=0.6544$meV, the dipolar interaction constant\cite{Aschroft} $U =
0.214727$meV\AA$^3$, the Land\'e $g$-factor $g=2$, the external
magnetic field ${\bv h}$ is fixed in the $z$-direction, and the
summation over $\left<i, j\right>$ is over all nearest neighbor
pairs. The dipolar interaction tensor ${\cal D}$ is given by:
\begin{equation}
\label{eq2}
{\cal D}_{ij}^{\alpha \beta}
= ( 3r_{ij}^\alpha r_{ij}^{\beta} - r^2_{ij}\delta_{\alpha \beta}) r_{ij}^{-5}.
\end{equation}
The ${\rm Mn}^{2+}$ ions are located on a body centered tetragonal lattice, with in-plane lattice constant
$a = 4.2$\AA, and $c$-axis lattice constant $c = 13.77$\AA. However,
it is known that the dipolar interaction between two tetragonal sublattices nearly vanishes due to the geometric arrangement of the moments.\cite{Lines67, Birgeneau70}
Therefore, besides a few simulations with two sublattices performed to check the validity of this assumption,
we included only one sublattice in most of our simulations,  which allowed us to simplify the dipolar summation
and to run simulations for larger systems.
Because the inter-layer interaction is weak, we have included
up to four layers of spins in our simulations, with open boundary condition in the $z$ direction.
Each layer is a square lattice with lattice constant equal to $a$
and the distance between adjacent layers equal to $c$.

The Hamiltonian  Eq.~(\ref{eq1}) is an approximation of the actual
quantum mechanical Hamiltonian, where spin operators have been
replaced with classical vector spins $S\bv{S}_i$ or
$\sqrt{S(S+1)}\bv{S}_i$. Here some ambiguities arise as to whether
$S$ or $\sqrt{S(S+1)}$ should be used. For the dipolar term, we
assume that the magnetic field generated by a spin is a dipole field
of a magnetic moment $gS\mu_B$, and the dipolar interaction energy
of a second spin with moment $gS\mu_B$ in this field is clearly
proportional to $S^2$.  This approximation guarantees that the total
dipolar energy of a ferromagnetic configuration agrees with
macroscopic classical magnetostatics of bulk materials. The exchange
term is more ambiguous. One can argue that $S(S+1)$ follows from the
quantum mechanical origin of the exchange interaction. After all,
the appropriate constant should reproduce the correct spin wave
spectrum or the critical temperature within acceptable error bars.
There is no guarantee that both of them can be accurately reproduced
with the same classical approximation. In general, by adopting the
classical approximation to spins, one admits an error possibly of
order $1/S$ in some quantities. To justify our choice in
Eq.~(\ref{eq1}), we first found that the critical temperature at
zero field of Eq.~(\ref{eq1}) was quite close to the experimental
value, then we turned on the magnetic fields to explore the full
phase diagram. It is unlikely that the entire experimental phase
diagram would be reproduced exactly including the spin-flop field.
However, our Monte Carlo simulations should exhibit the same
critical behavior as the real material, given that they are in the
same universality class. In particular, we want to test if there is
a ``real'' BCP at a finite temperature due to the long-range nature
of the dipolar interaction.

This paper is organized as the following: In Sec.~\ref{sec2}, we
briefly review the simulation techniques used in this research,
especially those designed to handle long-range, but very weak,
dipolar interaction; in Sec.~\ref{sec3}, we present the results from
simulations performed near each phase boundary; in Sec.~\ref{sec4}
we discuss the results and give our conclusions.

\section{Monte Carlo methods}
\label{sec2}
\subsection{Dipole summation}

Direct evaluation of the dipolar energy in Eq.~(\ref{eq1}) should be
avoided because the computational cost of direct evaluation scales
as $O(N^2)$ where $N$ is the number of spins, and the periodic
boundary condition needs to be satisfied. In our simulations we have
as many as  $8\times 10^4$ spins and need to evaluate the dipolar
energy repeatedly. Therefore, a fast algorithm for dipolar
interaction is required. We used the Benson and Mills
algorithm\cite{Benson69} which employs the fast Fourier
transformation of the spins to reduce the computational cost to
$O(N\ln N)$. After Fourier transform, the dipolar sum in
Eq.~(\ref{eq1}) can be written as
\begin{equation}
  \sum_{n,n',\alpha,\beta,\bv{q}} D_{n n'}^{\alpha\beta}(\bv{q}) S^\alpha_n(\bv{q}) S^\beta_{n'}(-\bv{q}),
\end{equation}
where $n$ and $n'$ label the different layers of the system, $\bv{q}$ is the in-plane wave vector,
and $D_{nn'}^{\alpha\beta}(\bv{q})$ is the Fourier transform of ${\cal D}_{i j}^{\alpha \beta}$.
This expression is less costly to evaluate than the Eq.~(\ref{eq2}), since the double summation over all the spins
is replaced by a single summation over the wave vectors, and $D_{nn'}^{\alpha\beta}(\bv{q})$ are constants which
can be calculated quickly in the initialization stage of the simulation. Explicit expressions for
$D_{nn'}^{\alpha\beta}(\bv{q})$ were first derived in Ref.~\onlinecite{Benson69}, and were reproduced in Ref.~\onlinecite{Filho00} with more detail and clarity.

\subsection{Monte Carlo updating scheme and histogram reweighting}

In Monte Carlo simulations of magnetic spin systems, cluster algorithms offer the benefit of reduced correlation times.
In Ref.~\onlinecite{Lee03}, the Wolff cluster algorithm\cite{Wolff89} was used to generate new spin
configurations based on the isotropic exchange term in the Hamiltonian. Although the Wolff algorithm is
rejection-free by itself, the new configuration
then has to be accepted or rejected with a Metropolis algorithm according to its dipolar and Zeeman energy.
The changes in the dipolar energy and Zeeman energy are roughly proportional to the size of the cluster generated
by the Wolff algorithm. When these changes are larger than $k_BT$, the number of rejections
rapidly increases, leading to substantially lower efficiency.
This problem occurs when the magnetic field is typically several Tesla in our simulations. On the other hand, in the paramagnetic phase or one of the ordered phases, the cluster size is small, the change
in dipolar energy is also small. It, thus, becomes redundant to evaluate the dipolar energy after every small change in the
spin configuration.

Since there are no rejection free algorithms for the dipolar
interaction, and the dipolar energy only contributes a fraction of
about 0.1 per cent to the total energy in our simulations, one of our
strategies to handle the dipolar interaction is to accumulate a
series of single spin flips before evaluating the dipolar energy,
then accept or reject this series of flips as a whole with the
Metropolis algorithm depending on the change of the dipolar energy.
The number of single spin flips for each Metropolis step can be
adjusted in the simulation so that the average acceptance ratio is
about 0.5, at which the Metropolis algorithm is most efficient. We
used the rejection-free heat-bath algorithm\cite{Miyatake86,
Loison04, Zhou04} to perform single spin flips, which handles both
the isotropic exchange and Zeeman terms in the Hamiltonian on the
same footing.

Although fast Fourier transform significantly reduces the
computational cost of dipolar interaction, this part is still the
bottle-neck of the simulation. Therefore, we want to further reduce
the number of dipolar energy evaluations. To this end, we separate a
short-range dipolar interaction from the full dipolar interaction.
The short-range part can be defined with an cutoff in distance. In
our simulations, we have included the up to fifth nearest in-plane
neighbor of each spin, and the spins directly above or below it in
the adjacent layer of the same sublattice, to form the short range
dipolar interaction. This short-range dipolar interaction can be
handled with the heat-bath algorithm on the same footing with the
exchange and the Zeeman term. The extra cost of evaluating local
fields produced by the additional 22 neighboring spins is
insignificant.  With this modification in single spin updates, the
Metropolis algorithm should be performed with respect to the change
in the long-range dipolar interaction, i.e., the difference between
the total dipolar energy and the short-range dipolar energy. Since
this long range dipolar energy is typically a small fraction (about
1 per cent) of the total dipolar energy, it is justified to accumulate many
single spin flips before refreshing the total dipolar energy.

We have found that the long-range dipolar energy in our simulations
is usually a fraction of about 0.001 per cent of the total energy, which is
actually comparable to $k_BT$. This allows us to further simplify
the above algorithm by removing the Metropolis step in the
simulation, while we simply calculate and record the full dipolar
energy for each configuration whose energies and magnetizations are
stored for histogram reweighting. In the end, we get a Markov chain
of configurations  from the simulation generated with a modified
Hamiltonian
\begin{equation}
{\cal H}' = {\cal H}_{\rm exchange} + {\cal H}_{\rm Zeeman} + {\cal H}_{\rm short},
\end{equation}
where the the first two terms are the exchange and Zeeman terms in
Eq.~(\ref{eq1}), and the last term is the short-range dipolar interaction.
For those configurations selected for computing thermodynamic averages,
we calculate and record ${\cal H}'$, ${\cal H}_{\rm short}$,
their full dipolar energy $H_{\rm dipole}$, staggered magnetization of each layer
\begin{equation}
    \bv{M}^\dagger_l = {1 \over L^2 } \sum_{i,j} (-1)^{i+j} \bv{S}_{ijl},
\end{equation}
where $L$ is the size of each layer and the index $l$ is the layer index,
and the average magnetization per spin in the $z$ direction
\begin{equation}
    M_z = { 1\over L^2 N_l} \sum_{i,j,l} S^z_{ijl},
\end{equation}
where $N_l$ is the number of layers in the system. As we have observed that
the interlayer coupling due to the dipolar interaction
is very weak, we define the total staggered magnetization $M^\dagger$ as
\begin{equation}
    M^\dagger =  \left[ N_l^{-1} \sum_l (M_l^\dagger)^2 \right]^{1/2}.
\end{equation}
Similarly, the Ising-like AF order parameter is defined as
\begin{equation}
\label{eq8}
    M^\dagger_z =  \left[ N_l^{-1} \sum_l (M_{l,z}^\dagger)^2\right]^{1/2},
\end{equation}
and the XY order parameter is defined as
\begin{equation}
\label{eq9}
    M^\dagger_{xy} =  \left[ N_l^{-1} \sum_l (M_{l,x}^\dagger)^2 + N_l^{-1} \sum_l (M_{l,y}^\dagger)^2 \right]^{1/2}.
\end{equation}
Note that we have ignored the factor $Sg\mu_B$ in the definitions of
various magnetizations so that they are normalized to 1 in the
antiferromagnetic configuration. Additionally, the fourth order
Binder cumulant for a quantity $Q$ is defined as
\begin{equation}
    U_4(Q) = 1- \frac{ \left< Q^4\right> }{3 \left<Q^2\right>^2},
\end{equation}
where $\left<\dots\right>$ represents the ensemble average.

The thermodynamic averages with respect to ${\cal H}'$ at a
temperature and a magnetic field slightly different from the
simulation can be obtained with the conventional histogram
reweighting technique.\cite{Swendsen88} To calculate the
thermodynamic average with respect to the original Hamiltonian, the
weight for each sample should be modified to
\begin{eqnarray}
\nonumber
&&\exp\left\{ - {1\over k_B T'} \left[{\cal H}' - Sg\mu_B M_z (h' - h) + {\cal H}_{\rm long} \right] \right\} \\
&&\times  \exp \left( { {\cal H}'  \over k_B T } \right),
\end{eqnarray}
where ${\cal H}_{\rm long} = {\cal H}_{\rm dipole}-{\cal H}_{\rm
short}$, $T$ and $h$ are the temperature and field at which the
simulation was performed, while $T'$ and $h'$ are the temperature
and field at which the histogram reweighting is done.

The performance of this perturbative reweighting scheme is valid
only when ${\cal H}_{\rm long}$ is smaller or comparable to the
thermal energy $k_B T$. For large system sizes, it has the same
problem as the conventional histogram reweighting methods, i.e., the
overlap of two ensembles defined by ${\cal H}$ and ${\cal H}'$
decreases exponentially, leading to a very low efficiency. In fact,
since both ${\cal H}_{\rm dipole}$ and ${\cal H}_{\rm short}$ are
extensive quantities, we expect their difference ${\cal H}_{\rm
long}$ to scale as $N_sL^2$. Therefore, it will exceed any given
$k_BT$  with a sufficiently large system size. For those large
systems, the above simulation scheme have to be modified to increase
the overlap between the two ensembles defined by ${\cal H}'$ and
${\cal H}$. Fortunately, even for our largest size $L=196$, the
long-range dipolar energy for a double layer system at about $T=20$K
and $h = 6$T is mostly positive around 4meV, and is mostly
distributed between $k_BT$ and $4k_BT$. Therefore, the perturbative
reweighting technique serves to increase the weight on those
configurations with lower dipolar energy, which are usually
associated with larger Ising order parameter. One might argue that
the long-range dipolar interaction could be ignored since it is
extremely small. Actually our simulations show that for the AF-PM
and XY-PM phase boundaries, the long-range dipolar interaction is
indeed negligible, but for the ``apparent" AF-XY phase boundary its
effect can be observed. With the perturbative reweighting technique,
we gain knowledge of both Hamiltonians, with or without long-range
dipolar interaction, simultaneously; hence we can tell where in the
phase diagram the long-range dipolar interaction changes the phase
boundaries.

Most of the results presented in the next section were calculated
with the perturbative reweighting technique, except part of the
results for the apparent spin-flop transition in Sec.~\ref{sec3c},
where a difference larger than the error bar is observed. For
equilibration, we ran two simulations from different initial
configurations until their staggered magnetizations converge within
statistical fluctuations. Then each simulation ran for $5\times
10^6$ to $2 \times 10^7$ Monte Carlo steps per spin to accumulate a
large amount of data for histogram reweighting. Early results for
zero field were compared with simulations with Metropolis
rejection/acceptance steps based on the full dipolar interaction; no
difference larger than the error bar had been observed.

\section{Results}
\label{sec3}
\subsection{Low-field antiferromagnetic transition}
\label{sec3a} 
\begin{figure}
\includegraphics[width=\columnwidth]{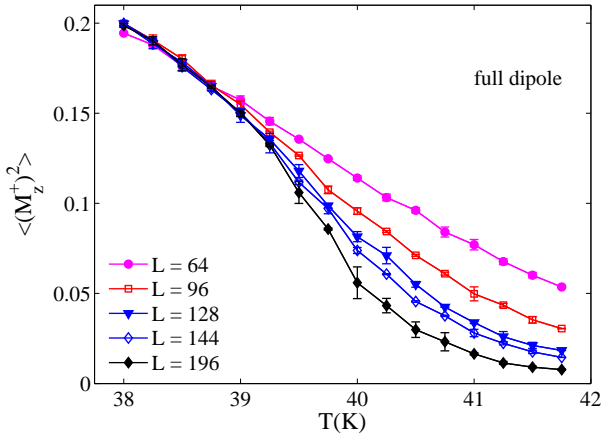}
\caption{(color online) Ising order parameter(staggered
magnetization)for double layer systems of different sizes across the
zero-field AF-PM phase transition. Data with full dipolar
interaction do not differ from those with only short range dipolar
interaction.} \label{fig1}
\end{figure}

\begin{figure}
\includegraphics[width=\columnwidth]{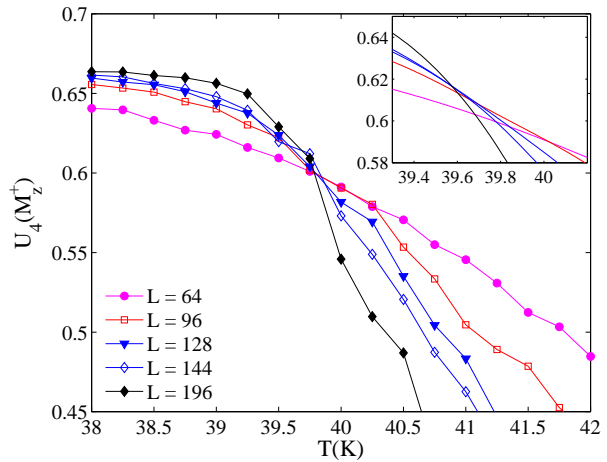}
\caption{(color online) The Binder cumulant for the Ising order
parameter across the AF-PM phase transition at zero field. The inset
shows a smooth spline fitting of the original data. Crossing points
in these curves approach the Ising universal value($\approx0.618$).}
\label{fig2}
\end{figure}

\begin{figure}
\includegraphics[width=\columnwidth]{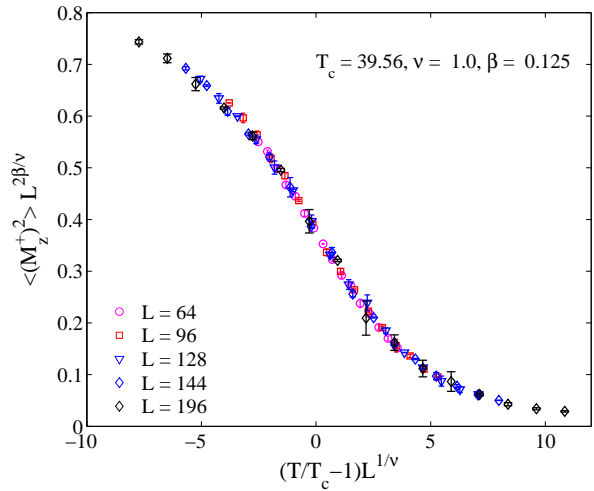}
\caption{(color online) Finite size scaling analysis of the AF-PM
phase transition at zero field. Data points are taken from
Fig.~\ref{fig1}. All of them fall onto a single curve with Ising
critical exponents.} \label{fig3}
\end{figure}

The zero-field AF-PM phase transition was studied with
Monte Carlo simulations in Ref.~\onlinecite{Lee03}, where $T_c$ (the
Neel temperature) was determined by extrapolating the crossing
points of the Binder cumulant. Since we have adopted a slightly
different model and also made a number of changes to the Monte Carlo
algorithm, we repeated this calculation for testing and calibration
purposes. The simulations were performed for double layer systems
with $L=64,96,128,144,196$. We also calculated the Binder cumulant
and performed finite size scaling analysis\cite{LandauBinder}  with
Ising critical exponents to fix the Neel temperature. Figure
\ref{fig1} shows the Ising order parameter (total staggered
magnetization in the $z$-direction) for different sizes at
temperatures close to the Neel temperature.
Although the Ising order parameter shows a strong size dependence
in the PM phase, the Neel temperature can not be determined directly
from it.  The Binder cumulant $U_4(M^\dagger_z)$ is plotted in
Fig.~\ref{fig2}. Unlike the results in Ref.~\onlinecite{Lee03},
where the crossing points of $U_4$ are above all 40K, we see in
Fig.~\ref{fig2} that all the crossing points are between 39.5K and
40K. The crossing points of these curves move up towards the
universal value of the Ising universality class ($U_4^* \approx
0.618$) as the system size increases. This trend is more clearly
revealed by curve fitting with smooth splines, shown in the inset of
Fig.~\ref{fig2}. Because data points for
$\left<(M_z^\dagger)^4\right>$ and $\left<(M_z^\dagger)^2\right>$
have smaller error bars, we actually did a curve fitting for those
two quantities first and plotted the Binder cumulant curve with the
fitted functions. $T_c$ can be fixed to be between 39.5K and 39.6K,
where the curves for three larger sizes cross. These observations
suggest that the critical behavior of this dipolar two-dimensional
Heisenberg antiferromagnet belongs to the Ising universality class.
Therefore, we performed a finite size scaling analysis to test this
prediction, as well as to fix the Neel temperature more accurately.
Figure~\ref{fig3} shows the finite size scaling analysis of the
Ising order parameter, where we plot  $(T/T_c-1)L^{1/\nu}$ versus
$\left<(M^\dagger_z)^2\right>L^{2\beta/\nu}$, with  Ising critical
exponents $\nu = 1$ and $\beta = 1/8$ .  Clearly, all the data from
different sizes fall nicely on a single curve. The best result is
achieved by choosing $T_c = 39.56$K. Obvious deviations from a
single curve are seen if $T_c$ changes by 0.1K, therefore we believe
the error bar for $T_c$ is less than $0.1$K.
Although we have obtained a $T_c$ which is only slightly smaller
than that obtained in Ref.~\onlinecite{Lee03}, our data for the
Ising order parameter and its Binder cumulant are noticeably
different from those in Ref.~\onlinecite{Lee03}. At the same
temperature, data presented here are smaller than those in
Ref.~\onlinecite{Lee03}. This difference is actually expected
because of the difference in the strength of the dipolar
interaction. The dipolar term is proportional to $S^2$ here in
Eq.~(\ref{eq1}), but proportional to $S(S+1)$ in the previous work.

\begin{figure}
\includegraphics[width=\columnwidth]{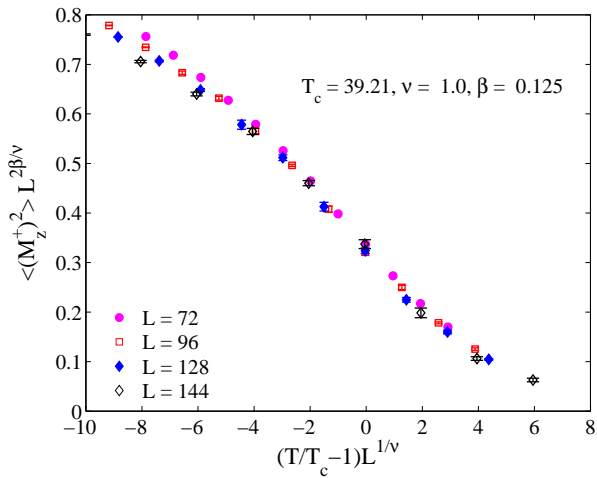}
\caption{(color online) Same scaling plot as Fig.~\ref{fig3}, but
for simulations performed at $h=3$T. The critical temperature, at
which the best collapsing of data points is achieved, is slightly
lower than that of Fig.~\ref{fig3}.} \label{fig4}
\end{figure}

We have also performed simulations at $h=3$T and 5T to study the
AF-PM phase transition in a finite magnetic field.  The
antiferromagnetic phase transition has been observed in both cases,
but the order parameter changes more gradually with temperature when
the magnetic field is turned on. Finite size scaling with Ising
exponents have been performed. Figure~\ref{fig4} shows the scaling
plot of $\left< (M_z)^2 \right>$ at $h = 3$T, which has a lightly
lower $T_c$. Long-range dipolar interaction only produces negligible
changes in these data points. The valid regime for finite size
scaling seems to be narrower than at $h=0$, because some deviations
are clearly seen in the low-temperature data points. This could be
due to the shape of the phase boundary, which is perpendicular to
the temperature axis at $h=0$ by symmetry, but not so at a finite
magnetic field. Because of this, we change both the temperature and
the effective anisotropy when the simulation scans temperature at a
constant magnetic field.

\subsection{Kosterlitz-Thouless transition }
\label{sec3b} 
\begin{figure}
\includegraphics[width=\columnwidth]{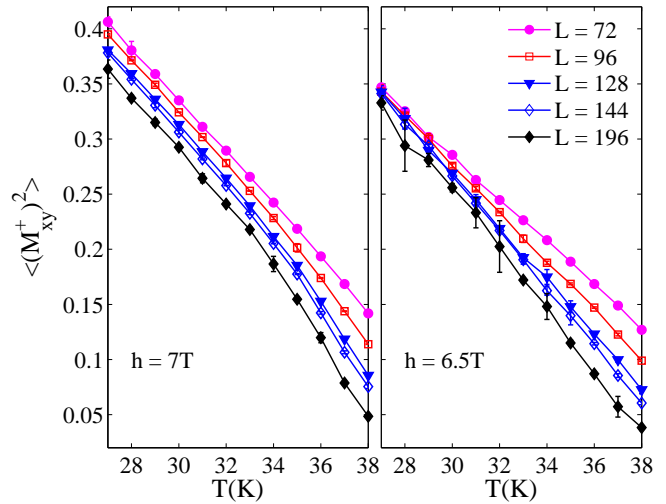}
\caption{(color online) Average XY order parameter across the XY-PM
phase boundary for double layer systems with different sizes.}
\label{fig5}
\end{figure}

When the magnetic field is above 6T, the AF-PM phase
transition disappears. Instead, the XY order parameter
Eq.~(\ref{eq9}) becomes large at low temperatures. For a
two-dimensional anisotropic Heisenberg antiferromagnet, one expects
to see an XY phase,\cite{Landau81, Leidl05, Zhou06} in which the
correlation function decreases algebraically. Since the dipolar
interaction breaks the spin rotational symmetry around the $z$ axis
on a square lattice, one would expect the XY phase to be destroyed
by its presence. In case of a ferromagnetic model, it has been shown
that above a critical strength, the ferromagnetic dipolar XY model
exhibits a ferromagnetic phase instead of an XY phase.\cite{Maier04}
Experimentally, a ``transverse'' phase with long-range order has
been found.\cite{Cowley93} However, since the XY phase is also very
sensitive to small perturbations such as crystal anisotropy and
disorder, it is not clear whether the dipolar interaction in ${\rm
Rb_2 Mn F_4}$ alone would prevent it from entering the XY phase. To
answer this question, we performed simulations in constant magnetic
fields $h=6.4, 6.5$ and 7T at temperatures from 27K to 38K.
Figure~\ref{fig5} shows the XY order parameter measured from these
simulations for double layer systems with $L=72,96,128,144$, and
196.
In all these simulations, the XY order parameter increases gradually
with lowering temperature in a broad range of temperature, and it is
hard to determine the transition temperature from Fig.~\ref{fig5}.
They also look very different from the results in
Ref.~\onlinecite{Leidl05}, where a transition in the XY order
parameter from zero to a finite value is clearly visible. There are
two reasons for this. First, the effective anisotropy induced by
dipolar interaction in ${\rm Rb_2MnF_4}$ is very weak. The dipolar
energy contributes only about 0.1 per cent to the total energy, while in
the anisotropic Heisenberg model studied in
Ref.~\onlinecite{Landau81,Leidl05,Zhou06}, the anisotropy is about
10 per cent to 20 per cent of the total energy (proportional to the anisotropy
constant $\Delta$). Secondly, the magnetic field at which the
simulations were performed (6.4T to 7T) is still close to the
apparent spin-flop transition at about 6.2T, where the system is
effectively an isotropic Heisenberg model. Experimentally, the
existence of such an effective Heisenberg model has been
tested.\cite{Christianson01} Near the apparent spin-flop transition,
the system has a large correlation length, which prevents the true
XY critical behavior from being revealed in simulations of limited
sizes. This also explains why in Fig.~\ref{fig5}
$\left<(M_{xy}^\dagger)^2\right>$ increases more rapidly at 7T with
decreasing temperature than it does at 6.5T.

\begin{figure}
\includegraphics[width=\columnwidth]{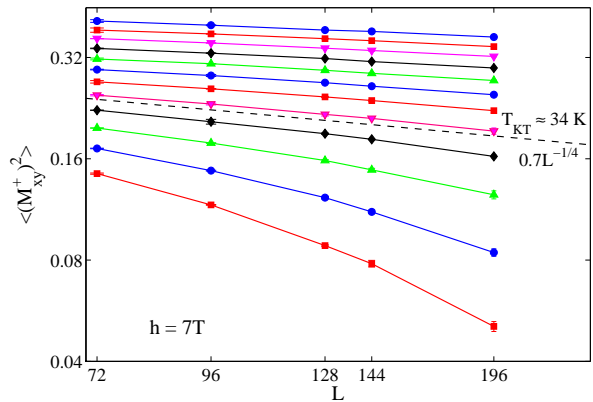}
\caption{(color online) Log-log plot of the size dependence of the XY order
parameter. The dashed line is a power-law with the critical exponent
$2\eta_c = 1/4$, used to identify the critical temperature. For each size,
the temperatures of the data points are 27K, 28K,$\dots$, 38K from top
to bottom. }
\label{fig6}
\end{figure}

Nevertheless, we can see in Fig.~\ref{fig5} that the XY order
parameter decreases with system size faster at higher temperatures
than at lower temperatures. In the PM phase, one expects the size
dependence to be exponential, i.e.,
$\left<(M_{xy}^\dagger)^2\right>\propto \exp(-2L/\xi)$; while in the
XY phase, the size dependence is power-law, i.e.,
$\left<(M_{xy}^\dagger)^2\right>\propto L^{-2\eta}$, where $\eta$ is
a temperature dependent exponent. On the XY-PM phase boundary, the
critical value of this exponent is $\eta_c = 1/8$. Therefore, we
plot $\left<(M_{xy}^\dagger)^2\right>$ versus $L$ in Fig.~\ref{fig6}
with log-log scale, and try to identify the critical temperature for
the Kosterlitz-Thouless transition.
Below the dashed line in Fig.~\ref{fig6},
the order parameter obviously decreases faster than
any power-law, which would be straight lines in the log-log scale.
Above it, the data points are very close to power-law, and their slopes
decrease with temperature. These features are consistent with an XY-PM phase
transition. The critical temperature $T_{KT}$ is roughly 34K, estimated from
Fig.~\ref{fig6}.
The same analysis has been done for simulations at $6.5$T and the estimated
$T_{KT}$ is also near 34K.

It has been found that if the square anisotropy is strong, the XY model confirms
the RG prediction that a second-order phase transition with nonuniversal critical exponents
occurs.\cite{Jose77, Regina04} If the anisotropy is weak, two possibilities for the phase diagram have been found
by Monte Carlo simulations:\cite{Rastelli04}
(1) a transition from the PM phase directly to the ferromagnetic phase, (2) a narrow
XY phase is sandwiched between the ferromagnetic phase and the PM phase.
Both of these cases might appear in our model if we replace the ferromagnetic phase
with an antiferromagnetic phase.
However, in all simulations performed above $h=6.4$T, at the lowest temperature $T=27$K,
we still see that the XY order parameter decreases with increasing system size.
No evidence for this phase is evident, at least for the range of lattice size that could be considered.
Based on this observation we believe if a low temperature
in-plane antiferromagnetic phase exists, it does not appear in the range of
temperature and magnetic field where our simulations have investigated.
Another check to exclude the transition from the PM phase to an Ising-like antiferromagnetic phase
is to do the finite size scaling
analysis with Ising exponents for the XY order parameter. We have found that
it is impossible to collapse all the data points in Fig.~\ref{fig5} onto
a single curve, no matter what critical temperature we use.

We have also performed simulations with a single layer of spins,
and the results  agreed with those for double layer systems within error bars.
The results without perturbative reweighting, i.e., short-range dipolar interaction
only, also do not differ noticeably from those with full dipolar interaction
presented in Fig.~\ref{fig5} and \ref{fig6}. Therefore, we conclude that our
results are consistent with an XY-PM transition. The main effect of the
dipolar interaction is to provide an easy axis anisotropy, but the
in-plane square anisotropy of the dipolar interaction is not strong enough to destroy the
XY phase in the parameter ranges that we have examined.

\subsection{The transition from AF phase to XY phase }
\label{sec3c}

Having found an Ising-like AF phase at low magnetic fields and
an XY phase at high magnetic fields, we now turn to the boundary
between these two phases. Precisely speaking, we want to tell if this
boundary exists in the thermodynamic limit, and if it exists, find where it
is connected to the XY-PM and AF-PM phase boundaries.  So far, we know
our system is best described by a two-dimensional anisotropic
Heisenberg antiferromagnet with a very weak long-range interaction of square
symmetry. Both the anisotropy and the long-range interaction come from
the dipolar interaction. If the long-range component of the dipolar interaction
can be completely ignored, the XY-PM phase boundary and the AF-PM
phase boundary meet at a zero-temperature BCP, as predicted by
RG theory\cite{Nelson76,Nelson77} and confirmed by Monte Carlo simulations recently.\cite{Zhou06}
In this case, there is no real phase boundary between the XY phase and the AF
phase. However, if the long-range component of the dipolar interaction is
relevant, then the other two possibilities might be favored, i.e., a BCP at a finite temperature
or a tetracritical point. In experiment, the neutron scattering data favored a finite temperature BCP,
so that the transition from the AF phase to the ``transverse'' phase is a first
order phase transition.\cite{Cowley93} Whatever brings the transverse phase,
which is observed to have long-range order, can also bring the bicritical
point to a finite temperature. Because both the transverse phase and the
AF phase have discrete symmetries, the BCP is not required
to have a continuous (rotational) symmetry. The existence of such a bicritical
point at finite temperature does not violate the Mermin-Wagner theorem.

We have performed simulations at constant temperatures $T=5,10,20$, and 30 K
and calculated both the Ising order parameter and the XY order parameter for
magnetic fields between 6T and 6.4T. We found that a transition apparently
occurs at about 6.2T at all temperatures, and this transition happens over
a larger range of magnetic field at higher temperatures than it does at lower
temperatures. It must be pointed out that the location of this transition
is about 0.9 to 1.1 T higher than the spin-flop transition in the experimental
phase diagram. The transition field also does not show a noticeable temperature
dependence, while the experimental spin-flop line has a positive slope.
However, our result is in agreement with previous simulations in Ref.~\onlinecite{Lee03},
therefore we believe this difference is a result of the classical approximation we
have adopted and also possibly some other weak effects, e.g., crystal field
anisotropy, that we have not included in our simulations.

\begin{figure}
\includegraphics[width=\columnwidth]{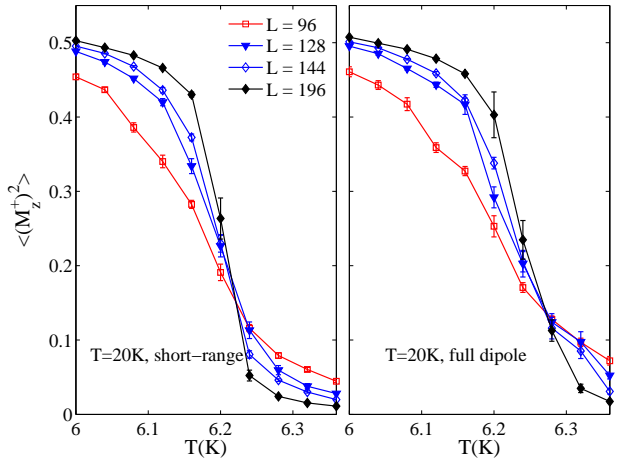}
\caption{(color online) Ising order parameter of double layer systems across the apparent spin-flop
transition at $T=20$K. The data reweighted with full dipolar interaction in the right panel shift
towards large magnetic field, and have larger error bars. }
\label{fig7}
\end{figure}

\begin{figure}[t]
\includegraphics[width=\columnwidth]{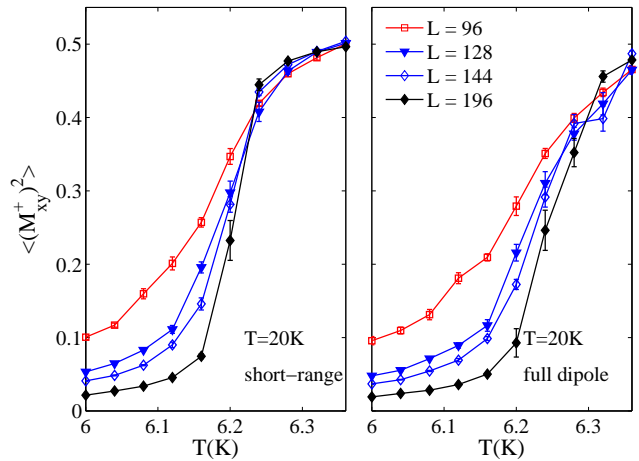}
\caption{(color online) XY order parameter of double layer systems across the apparent spin-flop
transition at $T=20$K. The data reweighted with full dipolar interaction in the right panel shift
towards large magnetic field, and have larger error bars. }
\label{fig8}
\end{figure}

\begin{figure}
\includegraphics[width=\columnwidth]{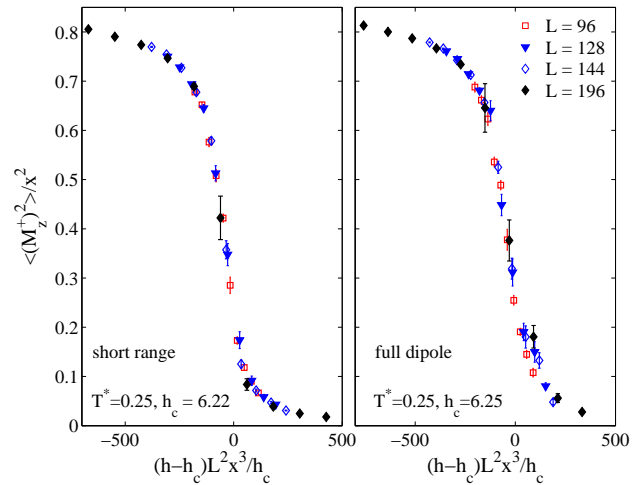}
\caption{(color online) Finite size scaling plot of the Ising order parameter at $T=20$K with
scaling ansatz for a zero-temperature BCP, where
$x=1-T^*\ln L /(2\pi)$}
\label{fig9}
\end{figure}

\begin{figure}
\centerline{
\includegraphics[width=\columnwidth, height=0.79\columnwidth]{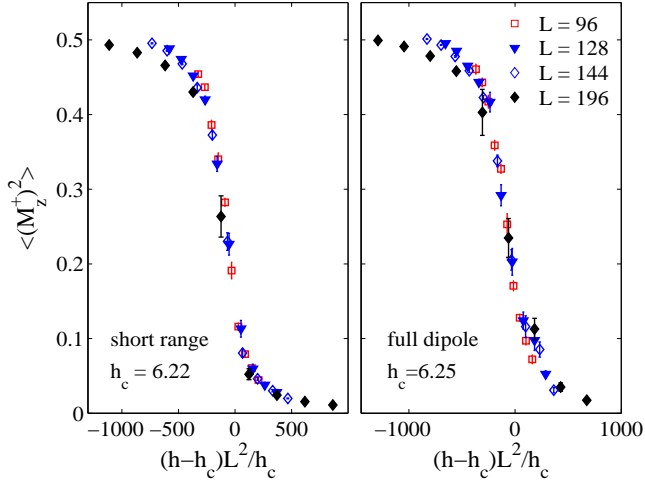}}
\caption{(color online) Finite size scaling plot of Ising order parameter at $T=20$K with
scaling ansatz for a first order phase transition, to compare with Fig.~\ref{fig9}. }

\label{fig10}
\end{figure}

Figure~\ref{fig7} shows the Ising order parameter calculated at $T=20$K
across the transition for different system sizes. The left panel shows
the result calculated with only short-range dipolar interaction, and the
right panel shows the same data reweighted with full dipolar interaction.
The XY order parameter which becomes large in higher
magnetic fields is shown in Fig.~\ref{fig8}.
To tell if there is a BCP at a finite temperature, we need to
classify the transition we have seen in Figs.~\ref{fig7} and \ref{fig8} using a
finite size scaling analysis. If it turns out to be a first order phase
transition, a BCP must exist above 20K. The finite size scaling
for the first order phase transition was established in
Ref.~\onlinecite{Binder84}.
For a BCP at $T=0$, Ref.~\onlinecite{Zhou06} showed that
logarithmic corrections to first order finite size scaling would be
observed. We plot the Ising order parameter  with the scaling
ansatz for the zero-temperature BCP \cite{Zhou06} in Fig.~\ref{fig9},
and with the first order scaling ansatz in Fig~\ref{fig10}.

In Fig.~\ref{fig9}, we have two tunable parameters: the critical field $h_c$ and
an effective temperature $T^*$. The logarithmic corrections, powers of
$x = 1-T^*\ln L/(2\pi)$, come from the spin renormalization constant calculated by
RG for an effective anisotropic non-linear $\sigma$ model at $T^*$, with
effective anisotropy vanishing at $h=h_c$.
By tuning $h_c$ and $T^*$, we have collapsed all the data points with
short-range dipolar interaction
onto a single curve very well. The data with full dipolar interaction also
collapse onto a single curve, except for a few data points with relatively large error bars.
Especially on the low-field side of the figure, the quality of collapsing is good.
On the other hand, the first order
scaling plot in Fig.~\ref{fig10} shows clear systematic deviation in the low-field
data points. This deviation is seen in both the left panel for short-range
dipolar interaction and the right panel for full dipolar interaction. The only
effect of the long-range part of the dipolar interaction is to shift the
critical field $h_c$ up by 0.03T. Although this effect is small, it is clearly out of
the error bars of the finite size scaling analysis. It is also expected from the
comparison of left and right panels in Figs.~\ref{fig7} and \ref{fig8}, where the
transition with the full dipolar interaction clearly shifts to higher magnetic fields.

\begin{figure}
\centerline{
\includegraphics[width=\columnwidth, height=0.77\columnwidth]{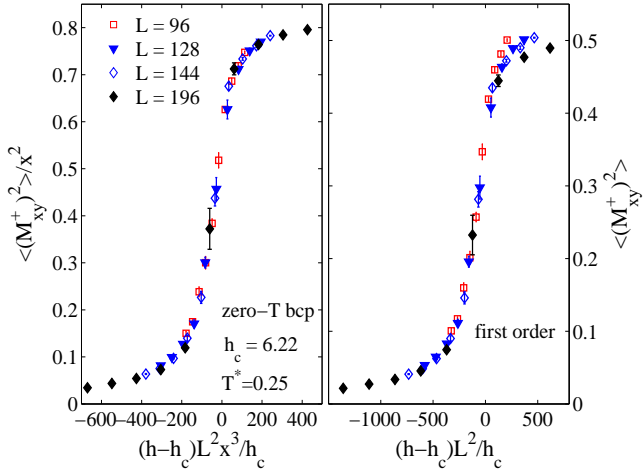}}
\caption{(color online) Finite size scaling of the XY order parameter at $T=20$K, comparison
of first order scenario and zero-temperature BCP. The critical
field $h_c$, and effective temperature $T^*$ are the same as those in Fig.~\ref{fig9}.}
\label{fig11}
\end{figure}

The same scaling analysis applies to the XY order parameters as well. Figure~\ref{fig11}
compares two finite size scaling plots for the XY order parameter at $T=20$K calculated
with short-range dipolar interaction. Obviously the scenario of a zero-temperature
BCP fits the data better than a first order phase transition.

\begin{figure}
\includegraphics[width=\columnwidth,height=0.77\columnwidth]{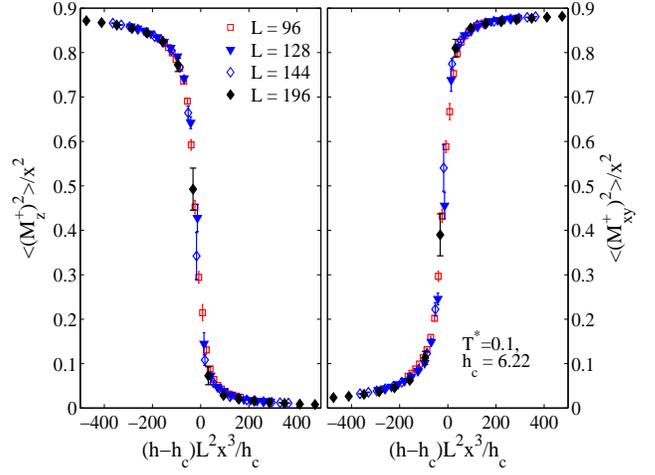}
\caption{(color online) Finite size scaling of the Ising (left) and XY (right) order parameter calculated
at 10K, corresponding to a zero-temperature BCP. Data shown here are calculated
with short-range dipolar interaction for double layer systems, data with histogram reweighting at different
magnetic are also shown. $h_c=6.22$ is the same as those in Fig.~\ref{fig9},
while $T^*=0.1$ is smaller here.}
\label{fig12}
\end{figure}

At lower temperatures, the same scaling behavior of order parameters has been observed,
and the critical field $h_c$ turns out to be nearly identical. Figure~\ref{fig12}
shows the finite size scaling plots for Ising and XY order parameter calculated at $T=10$K.
Since the transition at 10K happens within a narrower range of magnetic field, we have
included data points reweighted at fields different than that of the simulation. Data points for $L=196$
close to the transition which have large error bars are reweighted with different magnetic fields.
Nevertheless, most of the data points collapse nicely onto
a single curve. For data with short-range dipolar interactions, we have again found
$h_c = 6.22$T; while for data reweighted with full dipolar interaction, the scaling plots
look best if we choose $h_c=6.25$T.

\begin{figure}
\includegraphics[width=\columnwidth]{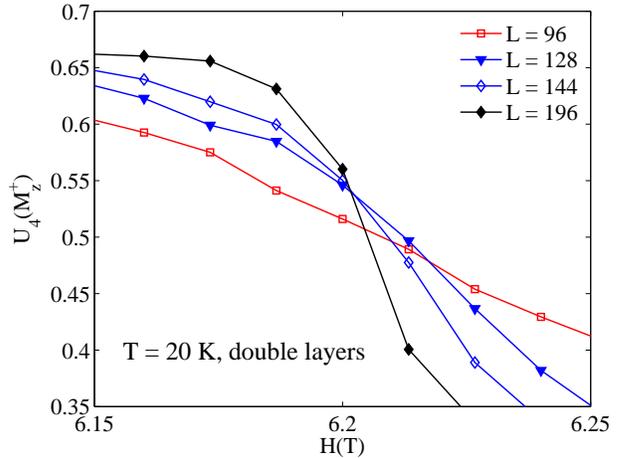}
\caption{(color online) The Binder cumulant of the Ising order parameter, three curves
for the larger sizes cross approximately at $h=6.203$T and $U_4 = 0.54$.}
\label{fig13}
\end{figure}

Therefore, our finite size scaling so far is more consistent with a zero-temperature
BCP than a finite temperature BCP above 20K.  Reference~\onlinecite{Zhou06} also predicts
finite size scaling relations for the susceptibility and specific heat, it also predicts
that the Binder cumulant $U_4(M^\dagger_z)$ is close to, but slightly below, 0.4 at the
critical field. We have observed the finite size scaling behavior of the susceptibility;
however we have not seen behaviors of the Binder cumulant and the specific heat similar
to those presented in Ref.~\onlinecite{Zhou06}. For the Binder cumulant, Fig.~\ref{fig13}
shows that the curves for three larger sizes cross approximately at $h=6.203$T and $U_4 = 0.54$.
This value is still very different from the universal value for the Ising universality class.
However, this is actually consistent with the theory in Ref.~\onlinecite{Zhou06}, if one notices
that here we have {\it two} nearly independent layers of spins.
If there is only one layer, Ref.~\onlinecite{Zhou06} has shown that at the critical field, the system is effectively
a single spin of length $\zeta$ with no anisotropy, where $\zeta$ is the spin
renormalization constant. Its angular distribution is uniform, which implies
$\left< (M_z^\dagger)^n\right> = 1/(n+1)$ and the crossing value of
$U_4(M_z^\dagger)$ is approximately 0.4.
In our simulations, since we have more than one layer, and they are weakly coupled,
we expect the total staggered magnetization of each layer $\bv{M}^\dagger_l$ is uniformly distributed on a sphere
of radius $\zeta$. Due to our definition of $M^\dagger_z$ in Eq.~(\ref{eq8}), the
distribution of $M^\dagger_z$ is not a uniform distribution, although $M^\dagger_{l,z}$
of each layer is distributed uniformly.  Suppose the interlayer coupling can be
completely ignored, which is a crude approximation. After some simple calculations,
we found the probability distribution of $s = (M^\dagger_z)^2/\zeta^2$ for a double
layer system is
\begin{equation}
P(s) = \left\{
\begin{array}{ll}
{\pi \over 2}, & 0< s \leq {1\over 2}, \\
\sin^{-1}{1\over \sqrt{2s}}-\sin^{-1}\sqrt{2s-1 \over 2s}, &1>s>{1\over 2}
\end{array}
\right. .
\end{equation}
Thus, if we ignore both the longitudinal fluctuation of staggered magnetization
and the interlayer coupling, the Binder cumulant at the critical field should be
 $1-\left<s^4\right>_P/(3\left<s^2\right>_P^2)$. A numerical evaluation of this
expression gives 0.5334, which is very close to the crossing point in Fig.~\ref{fig13}.
Therefore, our simulation is consistent with weakly coupled multiple layers of an anisotropic
Heisenberg antiferromagnet.

\begin{figure}
\includegraphics[width=\columnwidth]{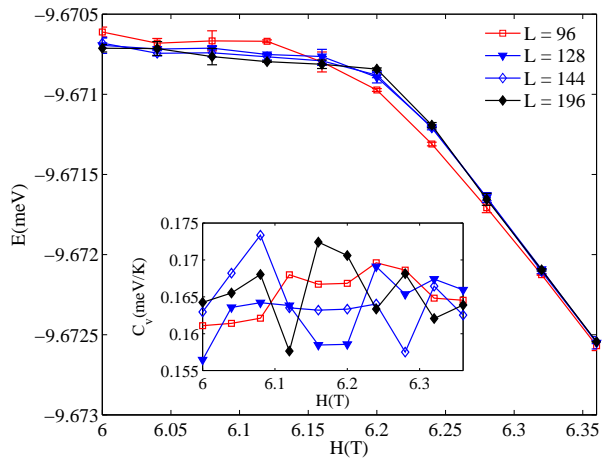}
\caption{(color online) The average energy per spin for a double layer system at $T=20$K
across the apparent spin flop transition. The inset shows the specific heat per spin, which
does not show a peak similar to that of first or second order phase transitions.
}
\label{fig14}
\end{figure}

As for the specific heat, we have not seen a peak at the transition in all our simulations.
Figure~\ref{fig14} shows the energy per spin and specific heat per spin calculated for double
layer systems at $T=20$K with short range dipolar interaction. The energy drops when the
magnetic field is larger than the critical field. However the specific heat shown in the
inset does not show any sign of a peak. Although the error bar of the specific heat,
as one can estimate from the fluctuation of the data points,  is about 10 per cent,
a peak which is expected to be similar to those discovered
in Ref.~\onlinecite{Zhou06},  is clearly absent.
However, this result is actually consistent with the finite size scaling theory for specific
heat in Ref.~\onlinecite{Zhou06}, which shows that the peak in specific heat should
be proportional to $(dh_c /d T)^2$. Because the critical field of our model
is almost independent of the temperature, i.e., $dh_c /d T \approx 0$, we actually
do not expect to see a peak in the specific heat here.

\subsection{ Discussions }
\label{sec3d}

\begin{figure}[t]
\includegraphics[width=0.6\columnwidth]{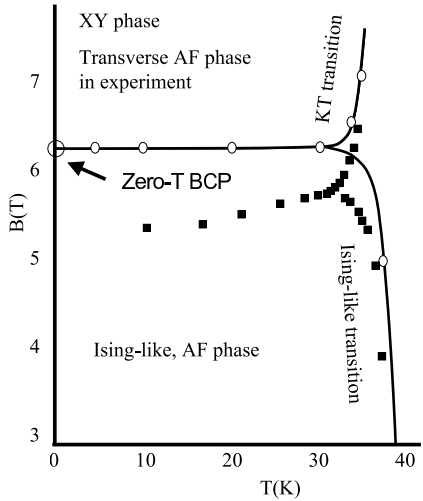}
\caption{Comparison between our phase diagram and the experimental
results. The experimental data points from
Ref.~\onlinecite{Cowley93} are ploted in solid squares.}
\label{fig15}
\end{figure}

To summarize our results, we construct a phase diagram in
Fig.~\ref{fig15} based on our simulations and compare it to the
experimental phase diagram from Ref.~\onlinecite{Cowley93}. Both our
XY-PM and AF-PM phase boundaries are close to experimental results,
the most pronounced difference is the spin-flop line. Rigorously
speaking, our spin-flop line is not a single line, but the
extensions of XY-PM and AF-PM phase boundaries which are
exponentially close to each other and meet at a zero-temperature
BCP. The experimental XY-AF ``phase boundary" is empirical. Our
spin-flop line is higher in magnetic field than the experimental one
and has a nearly vanishing slope, but this difference in spin-flop
field is most likely to be a consequence of the classical
approximation which omitted quantum fluctuations of the spins. The
anisotropic Heisenberg antiferromagnet studied in
Ref.~\onlinecite{Zhou06} offers an simple case to qualitatively
analyze this effect. A brief derivation of the spin-flop field of
this model is given in the appendix. If we assume the length of the
classical spins is $\sqrt{S(S+1)}$, the zero-temperature spin-flop
field of this simple model in the classical case is
$4J\sqrt{S(S+1)(1-\Delta^2)}$. The spin-flop field of the quantum
mechanical Hamiltonian is found to be $4JS\sqrt{1-\Delta^2}$ within
the linear spin-wave approximation. More accurate results can be
obtained by quantum Monte Carlo simulations, however, the linear
spin-wave theory has already considerably reduced the spin-flop
field. Since this simple model and the dipolar Heisenberg
antiferromagnet studied here have the same critical behavior near
the apparent spin-flop transition, one would also expect the quantum
effects in the latter model would reduce the spin-flop field by
approximately the same amount. Acutally, given the classical result
$h_c \approx 6.25$T, assuming the classical model consists of spins
of length $\sqrt{S(S+1)}$, the reduced spin-flop transition would be
$h_c/\sqrt{1+1/S} = 5.28$T, which happens to be in agreement with
the experimental value.

Above the spin-flop line, we have observed the XY phase, as far as
our simulations have covered, while the experiment shows a transverse phase. Therefore, our
Hamiltonian certainly misses some weak but important effects in the real material, as the
intricate correlation of the XY phase and the spin-flop transition is sensitive to many
perturbations. Disorder is one of them, which can impose a cutoff in correlation length
of the system so that the system would not approach the ideal zero-point BCP from the narrow
PM phase. As a result, an apparent finite temperature BCP would be observed and the apparent
spin-flop transition below the ``BCP" looks like a first order transition. The disorder can come from both
the crystal defects and slight inhomogeneity in the magnetic field. The experimentally
observed finite temperature BCP can also be a result of crossover to three dimensions due to
very weak exchange between layers.

The other facter that might have contributed to a phase diagram
different from the experimental result is the exchange constant. The
spin-wave analysis of $\rm Rb_2 Mn F_4$, which provided us the
exchange constant $J$,
 were done for systems in zero magnetic field, and the dipolar interaction had already been
simplified to a temperature dependent staggered magnetic field acting on Mn$^{2+}$ spins.\cite{Wijin73}
Therefore, the exchange integral provided by this theory is an effective quantity that depends on the
particular form of the Hamiltonian which has been assumed.
As far as we know, similar calculations have not been done in magnetic fields close to the
spin-flop transition. It is not guaranteed that when the full dipolar interaction is used in the Hamiltonian,
instead of an effective staggered magnetic field, the exchange integral
deduced from a simplified Hamiltonian is still applicable and can be treated as a constant
independent on either temperature or magnetic field.

\begin{figure}
\includegraphics[width=\columnwidth]{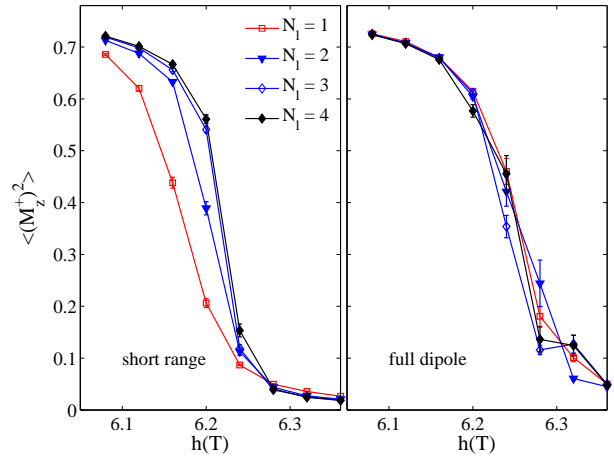}
\caption{(color online) Ising order parameters calculated for
systems at $T=10$K with $L=96$ and different number of layers. The
thickness dependence is weak. It is more obvious in the left panel 
where we only include short-range dipolar interaction, than the right
panel with full dipole reweighting. }
\label{fig16}
\end{figure}

\begin{figure}
\includegraphics[width=\columnwidth, height=0.8\columnwidth]{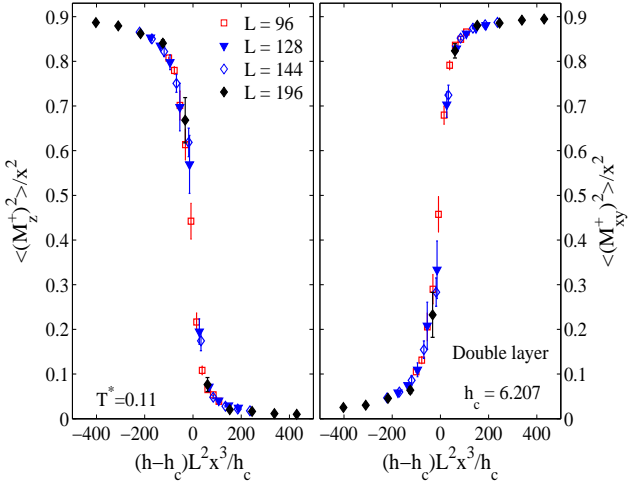}
\caption{(color online) Finite-size scaling plot for simulations at
$T=10$K with inter-sublattice dipolar interactions. The data in this
figure only differ very slightly from those in Fig.~\ref{fig12}, in
which the inter-sublattice dipolar interactions were omitted. }
\label{fig17}
\end{figure}

Finally, we show some results that justify two main assumptions,
i.e., the inclusion of only a few layers of Mn$^{2+}$ spins, and the
omission of two sublattices.  Figure~\ref{fig16} shows the Ising order
parameter across the apparent spin-flop transition for systems with
$L=96$ but different number of layers. With short-range dipole
interaction, the result seems to saturate when we have three or more
layers. After reweighting with full dipolar interaction, the
difference between data for different number of layers becomes even
smaller. We estimate the change in $h_c$ due to the change in number
of layers should be of order 0.01T. Therefore, it is justified to do
simulations with only a few layers of spins. The crossover to a
three dimensional system will only occur at very low temperatures.
Figure \ref{fig17}shows a finite-size scaling plot of  the apparent
spin-flop transition at $T=10$K calculated with two sublattices. The
dipolar interactions between two sublattices were truncated to third
nearest neighbors, i.e., an Mn$^{2+}$ spin feels the magnetic field
generated by totally 32 neighboring spins in the Mn$^{2+}$ layer
above and below it belonging to the other sublattice. The magnetic
field contributed by spins outside this truncation radius should be
extremely small based on our experience with the long-range dipolar
interaction. Compared with Fig.~\ref{fig12}, which was calculated
with a single sublattice, the difference in $T^*$ and $h_c$ is
negligible. We have enough reason not to expect the interaction
between two sublattices to reduce the apparent spin-flop field $h_c$
by more than $0.1$T. The actual additional energy due to the
inter-sublattice dipolar interaction is found to be only comparable
to the long-range dipolar energy.

\section{Conclusions}
\label{sec4} In conclusion, we have tried to explain the phase
diagram of $\rm Rb_2 Mn F_4$ using a classical spin model with
dipolar interactions. A large amount of Monte Carlo simulations have
been carried out to investigate the phase boundaries. Among
different strategies to handle the dipolar interaction in the
simulations, we have found our perturbative reweighting technique to
be  the most suitable for very weak dipolar interactions in $\rm
Rb_2 Mn F_4$. The phase diagram inferred from our data captures the
main features of the experimental phase diagram and the agreement is
good at low magnetic fields. On the apparent spin-flop line, the XY
and AF boundaries come so close together that they cannot be
distinguished below an ``effective" BCP at $T\approx 30$K. However,
our data analyses support a zero temperature BCP. This conclusion is
based on a novel finite size scaling analysis for two-dimensional
anisotropic Heisenberg antiferromagnets.\cite{Zhou06} If this
multicritical point is located at very low finite temperature, as
suggested by Ref.~\onlinecite{Vicari07}. We believe its temperature
must be sufficiently low, which is beyond our numerical accuracy.
The ground state degeneracy for the anisotropic Heisenberg
antiferromagnets as found in Ref.~\onlinecite{Selke07} may also
exist in our model with dipolar interactions, which we have not yet
verified. If it exists, one might simply rename the bicritical point
as a tetracritical point. The zero temperature BCP is located above
the experimental spin-flop line in the phase diagram, which appears
to be a a line of first order phase transitions. We believe this
difference from the experimental phase diagram is mainly caused by
the classical approximation.  Nevertheless, we have confirmed that
the dominant effect of the dipolar interaction in  $\rm Rb_2 Mn F_4$
is to provide an effective anisotropy, while other effects, such as
in-plane square anisotropy and interlayer interaction, are extremely
weak. Therefore, we would hope to obtain a more accurate phase
diagram if we performed quantum Monte Carlo simulations for a
simpler Hamiltonian which includes the effective anisotropy.

\begin{acknowledgments}
We thank W. Selke, E. Vicari, and A. Pelissetto for fruitful
discussions. This research was conducted at the Center for Nanophase
Materials Sciences, which is sponsored at Oak Ridge National
Laboratory by the Division of Scientific User Facilities, U.S.
Department of Energy.

\end{acknowledgments}

\appendix*
\section{ Spin-flop field at $T=0$ of anisotropic Heisenberg antiferromagnet}
As an analogy to the dipolar Heisenberg antiferromagnet, we consider the simple
anisotropic Heisenberg antiferromagnet with the Hamiltonian:
\begin{equation}
\label{eqa1}
{\cal H}  = J \sum_{\left< i,j\right>}
\left[\Delta\left( S_i^xS_j^x +  S_i^yS_j^y \right) + S_i^zS_j^z\right].
-H\sum_i S_i^z,
 \end{equation}
which is defined on a square lattice. The classical version of this
model has been well studied,\cite{Landau81, Leidl05, Zhou06} where
the spins are treated as unit vectors. The spin-flop field at
zero-temperature is $H_c = 4J \sqrt{1-\Delta^2}$. If we replace the
spins with vectors of length $\sqrt{S(S+1)}$, $H_c$ is then modified
to $4J\sqrt{S(S+1)(1-\Delta^2)}$. This is not the only way to make
connections to the quantum Hamiltonian. One can also replace $J$
with $JS(S+1)$, while replacing $H$ with $HS$, which can be
justified by arguing that the Zeeman energy of the ferromagnetic
configuration takes on the correct macroscopic value. In this case,
the spin-flop field  is modified to $H_c = 4J(S+1) \sqrt{1-\Delta^2}
$. However, in any case, we will show that the classical spin-flop
field is larger than the quantum mechanical spin-flop field. By
introducing the Holstein-Primakoff (HP) bosons on $A$ and $B$
sublattices respectively, and keeping the quadratic terms, the
Hamiltonian Eq.~(\ref{eqa1}) can be rewritten as
\begin{eqnarray}
\nonumber
{\cal H} &\approx& J \sum_{i\in A }\sum_{\left<j,i\right>} \left[
  \Delta S \left( \ad_i \bd_j  + {\rm c.c.} \right)
+ (\ad_i a_i -S)(S-\bd_j b_j)
\right] \\
&&- H \sum_{i \in A}\ad_i a + H \sum_{j \in B}\bd_i b ,
\end{eqnarray}
where $a$ and $\ad$ are HP boson operators on sublattice $A$, $b$
and $\bd$ on sublattice $B$, index $i$ labels sites on sublattice
$A$ which are nearest neighbors of the sites on sublattice $B$
labeled with $j$. After a Fourier transformation, this quadratic
Hamiltonian turns out to be
\begin{equation}
{\cal H} = -4JS(S+1)N_A - HN_A - \sum_{\bv{k}} {\cal H}_\bv{k},
\end{equation}
where $N_A$ is the number of sites on sublattice A, and
\begin{equation}
{\cal H}_\bv{k} = SJ
  \left(
  \begin{array}{ll}
  \ad_\bv{k} & b_{-\bv{k}}
  \end{array}
  \right)
  \left(
 \begin{array}{ll}
 4-h & \Delta \gamma_{\bv k} \\
 \Delta  \gamma_{\bv k} & 4+h
  \end{array}
  \right)
  \left(
   \begin{array} {l}
   a_\bv{k} \\
   \bd_{-\bv{k}}
  \end{array}
  \right).
\end{equation}
For simplicity, we have defined $h = H/SJ$ and $\gamma_{\bf k} = 2\cos k_x + 2\cos k_y$.
The spin-wave spectrum can be obtained with the Bogoliubov transformation:
\begin{eqnarray}
c_{\bf k} = \cosh \theta_{\bf k} a_{\bf k} + \sinh \theta_{\bf k} \bd_{-\bf k}, \\
d_{\bf k} = \sinh \theta_{\bf k} \ad_{\bf k} + \cosh \theta_{\bf k} b_{-\bf k}.
\end{eqnarray}
In order to eliminate the cross terms in the Hamiltonian,
one sets $\tanh 2\theta_{\bf k} =\Delta \gamma_{\bf k} /4$.
Apart from a constant term,  the spin-wave part of the Hamiltonian turns out to be
\begin{equation}
{\cal H} _{\rm sw} =  \sum_\bv{k} \left[ \omega_+(\bv{k}) \dd_\bv{k}
d_\bv{k}+\omega_-(\bv{k})\cd_\bv{k} c_\bv{k}\right],
\end{equation}
where
\begin{equation}
\omega_\pm(\bv{k})  =  JS\sqrt{16 - \Delta^2 \gamma_{\bf k}^2 } \pm H.
\end{equation}

When $H$ is large enough such that $\omega_-(0)$ becomes negative,
the AF ground state becomes unstable since the excitations on
spin-wave mode $c_{\bv{k}=0}$ lower the ground state energy. This
precisely indicates the the spin-flop instability. Therefore, the
critical magnetic field is given by
\begin{equation}
H_c = 4JS \sqrt{1-\Delta^2 }.
\end{equation}
Although the above spin-wave analysis is only a crude approximation,
we see that the quantum effect lowers the spin-flop field by a
factor of $S/(S+1)$ or $\sqrt{S/(S+1)}$, depending on which
classical approximation one uses. The case with $\Delta = 2/3$ and
$S=1/2$ has been studied with quantum Monte Carlo
simulations.\cite{Schmid02} Its phase diagram shows the spin-flop
field is at approximately $h/J_{xy} = 1.8$, i.e., $H_c = 1.2J$ in our
notation here. The above spin-wave approximation gives $H_c = 1.49
J$, and the two classical approximations gives $H_c = 4.47J$ and
$H_c = 2.58J$ respectively. Clearly, the classical approximations
overestimate the spin-flop field. The difference from the real
spin-flop field is large as we expect the quantum fluctuation to
have a strong effect for $S=1/2$. For larger spins, such as $S=5/2$
which is studied in this paper, the classical approximation should
work better. However, we still expect it to overestimate the
spin-flop field by an noticeable amount.

\bibliographystyle{apsrev}
\bibliography{article}
\end{document}